# A new semi-supervised self-training method for lung cancer prediction

Kelvin Shak[a], Mundher Al-Shabi[a], Andrea Liew[a], Boon Leong Lan[a], Wai Yee Chan[b], Kwan Hoong Ng[b], Maxine Tan[a,c]

a) Electrical and Computer Systems Engineering and Advanced Engineering Platform, School of Engineering, Monash University Malaysia, Bandar Sunway 47500, Malaysia
b) Department of Biomedical Imaging, University of Malaya, 50603 Kuala Lumpur, Malaysia
c) School of Electrical and Computer Engineering, The University of Oklahoma, Norman, OK 73019, USA

Corresponding author: Kelvin Shak (Kelvin.Shak@monash.edu)

## Abstract

**Background and Objective**: Early detection of lung cancer is crucial as it has high mortality rate with patients commonly present with the disease at stage 3 and above. Although many automated schemes have been developed, there are only relatively few methods that simultaneously detect and classify nodules from computed tomography (CT) scans. Furthermore, very few studies have used semi-supervised learning for lung cancer prediction. This study presents a complete end-to-end scheme to detect and classify lung nodules using the state-of-the-art Self-training with Noisy Student method on a comprehensive CT lung screening dataset of around 4,000 CT scans.

**Methods**: We used three datasets, namely LUNA16, LIDC and NLST, for this study. We first utilise a three-dimensional deep convolutional neural network model to detect lung nodules in the detection stage. Then, we developed a new classification model called Maxout Local-Global Network to accurately predict large variations of nodules. The model uses non-local networks to detect global features including shape features, residual blocks to detect local features including nodule texture, and a Maxout layer to detect nodule variations. We also trained the first Self-training with Noisy Student model to predict lung cancer on the unlabelled NLST datasets. Then, we performed Mixup regularization to enhance our scheme and provide robustness to erroneous labels. We performed extensive ablation studies to analyze each component of our new scheme.

**Results**: Our new Mixup Maxout Local-Global network achieves an AUC of 0.87 on 2,005 completely independent testing scans from the NLST dataset. This is the highest achieved AUC result on the highest number of CT scans in the literature. Our new scheme significantly outperformed the next highest performing method at the 5% significance level using DeLong's test ($p = 0.0001$).

**Conclusions**: This study presents a new complete end-to-end scheme to predict lung cancer using Self-training with Noisy Student combined with Mixup regularization. On a completely independent dataset of 2,005 scans, we achieved state-of-the-art performance even with more images as compared to other methods.

# 1. Introduction

The prevalence of lung cancer is increasing yearly and with the highest mortality rate among other types of cancer [1]. It is estimated that 228,820 new lung cancer cases will be diagnosed in the year 2020 in the United States [2]. With the advance of medical and surgical treatment shifting towards precision medicine, early detection of the disease will significantly improve patients' survival rate.

To date, most researchers have successfully developed methods to detect or classify lung nodules independently. However, there are only relatively few methods that can simultaneously detect *and* classify nodules from computed tomography (CT) scans [3]. To address this shortcoming, we present in this study an end-to-end scheme to detect and classify lung nodules in order to predict lung cancer, which is evaluated on a comprehensive CT lung screening dataset of around 4,000 CT scans.

A general difficulty in the medical imaging field is the unavailability of sufficient labelled/annotated data to validate new deep learning methods [4]. Relatively small annotated datasets including LUNA16 and LIDC-IDRI [5,6] are typically used for lung nodule detection or classification. The LUNA16 and LIDC-IDRI datasets contain detailed annotated locations of each lung nodule within the CT scans. The National Lung Screening Trial (NLST) lung screening dataset [7], released by the National Cancer Institute (NCI), is a vast resource of lung screening CT scans; however, a downside of the NLST dataset is that the ground truth data only provides a final diagnosis of cancer (i.e., 1) or cancer-free (i.e., 0) for each patient and does not provide the labelled/annotated locations of the nodules within the CT scans.

To overcome this deficiency in the NLST dataset, we use Self-training with Noisy Student Training [8] to leverage on the unlabelled data/nodules - to the best of our knowledge this has not been done before. Self-training has produced state-of-the-art results on image classification in ImageNet [8]. Using the self-training method, a trained teacher model will generate pseudo labels on unlabelled images. Then, a student model is trained on both labelled and pseudo-labelled images, and noise is simultaneously added to improve the student's learning capacity. In our self-training framework, we use our state-of-the-art Local-Global network [9] to classify unlabelled nodules in the NLST dataset and train a new student model with noise on the pseudo-labelled data. Similar to the results obtained in the original self-training paper [8], we obtained significant improvement in our lung nodule classification scheme in terms of the area under the receiver operating characteristic curve (AUC) and we obtained state-of-the-art results for lung cancer prediction on the NLST dataset.

Our main contributions are as follows:

1. We develop a new automatic end-to-end scheme that simultaneously detects and classifies lung nodules to predict lung cancer from CT scans.
2. We implement the first Self-training with Noisy Student model for lung cancer prediction on the unlabelled NLST dataset and show that the results generalize to big independent datasets.
3. We enhance our state-of-the-art lung nodule classification scheme, namely the Local-Global network, by implementing the Mixup technique to improve final model performance and including a Maxout Layer to classify large variations of lung nodules.
4. We train and develop our new scheme on around 4,000 CT scans and achieve state-of-the-art performance even with more images as compared to other studies.

## 2. Related work

Many deep learning based detection models are available in the literature including DeepLung, DeepSEED, NODULe, 3D semi-supervised convolutional transfer neural network and Lung Nodule detection using Faster R-CNN [10–14]. Recently, deep learning models have also been shown to differentiate between malignant and benign nodules very well [9,15–17]. Although deep learning has achieved state-of-the-art performance in many fields, considerable effort is still invested to enhance deep learning's performance. More recently, efforts to do so have focused on semi-supervised and unsupervised learning [8,18], with some work in the medical imaging field [12,19].

Self-training has emerged as one of the highest performing and state-of-the-art methods in this field [8]. With self-training, a pre-trained teacher model is used to train a student model. By adding noise to the student model and training it on pseudo-labelled datasets, the performance of the student model improves and outperforms the teacher model.

Semi-supervised learning is another popular method that is related to self-training and is widely researched on. For example, ATLASS, which is an open-source tool based on medical imaging, has combined transfer learning with semi-supervised learning to train models with very few annotated images available [20]. Some of the other models that utilize semi-supervised learning are used for segmentations in MRI and ultrasound images. Burton et al. [21] implemented semi-supervised learning for automated segmentation of knees that achieved results on par with models that were trained with labelled data [22–24]. Similarly, Han et. al performed segmentation using semi-supervised learning, but on lesions from breast ultrasound images, where 1900 out of the 2000 training images were unannotated images [25]. This demonstrates the ability of semi-supervised methods in leveraging on unlabelled medical images.

## 3. Methods

In the following subsections, we describe our new automatic end-to-end scheme depicted in Figure 1 to simultaneously detect and classify lung nodules to predict lung cancer in CT scans.

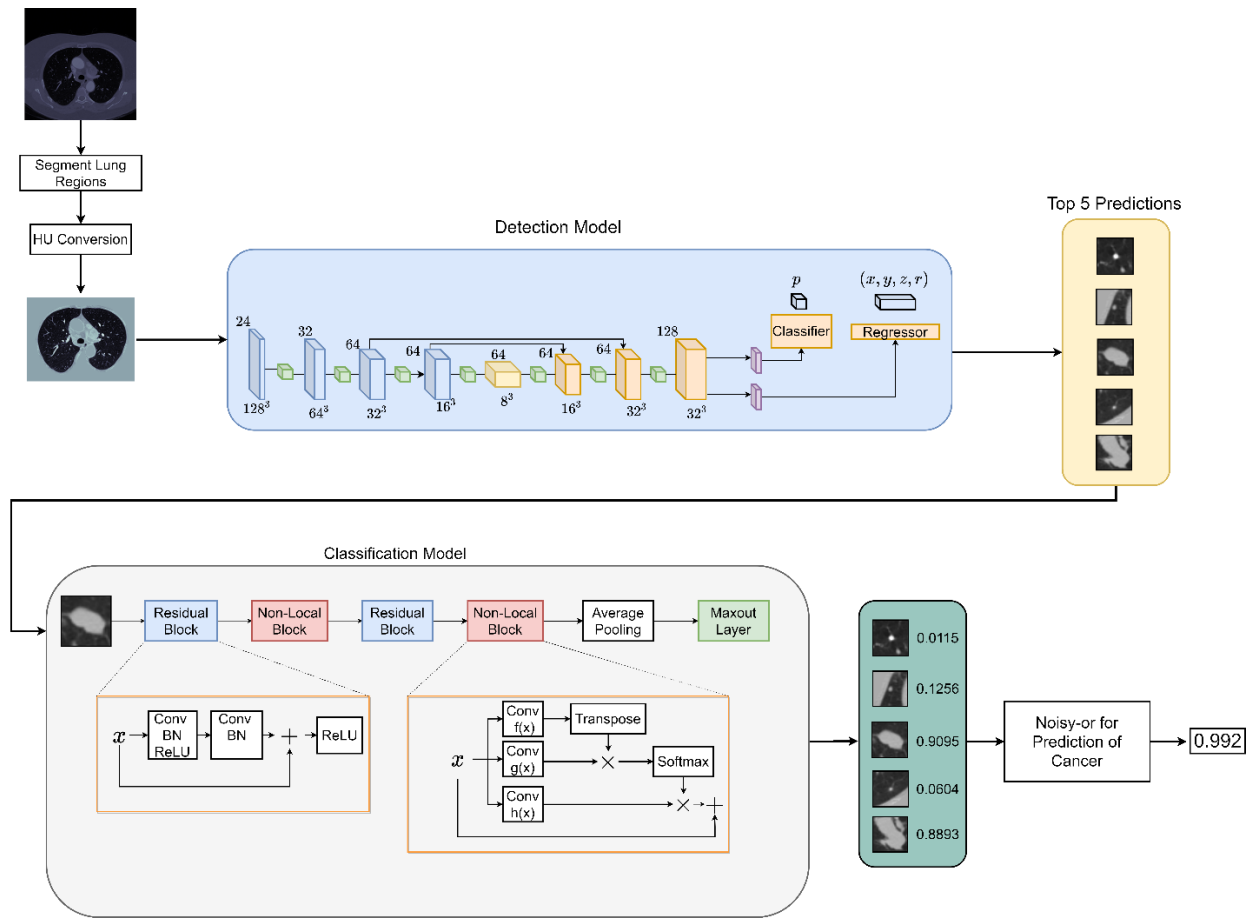

*Fig. 1. Block diagram of our new end-to-end scheme to predict lung cancer in patients. Conv stands for convolution; BN stands for batch normalization and ReLU is the implemented activation function.*

### 3.1 Nodule detection network

Our nodule detection method is based on DeepSEED's detection model [11]. DeepSEED's model is based on a novel 3D CNN architecture and utilises an encoder-decoder structure with the implementation of ResNet blocks. The authors in [11] built a Region Proposal Network (RPN) upon the encoder-decoder structure and incorporated the Squeeze-and-Excitation [26] architecture to learn image features.

In our experiments, we retrained the model from scratch using the LUNA16 [27] dataset and modified the batch size of the original DeepSEED model. We achieved the highest sensitivity result of 0.893 on our modified model. We ran the modified DeepSEED model on the NLST datasets to generate predictions for them. We selected the top 5 predictions and fed them to the next stage of our scheme for benign/malignant nodule classification.

### 3.2 Nodule classification network

The next stage is to classify the detected nodules as benign/malignant. To achieve this, we modified our state-of-the-art nodule classification scheme called Local-Global Network [9] that implements a non-local neural network introduced by Wang et al. [28], which is also known as the self-attention layer. In our Local-Global Network architecture, we implemented the non-local network with CNNs. The non-local network enables us to extract global features, such as nodule shape and boundary information (e.g., spicularity), whereas the CNN extracts local features, such as nodule texture. Thus, combining CNN layers with non-local layers enables us to extract both local and global features, which enables us to classify lung nodules more accurately.

In the original Local-Global Network architecture, we implemented a linear layer before the sigmoid activation function [9]. In this study, we implement a Maxout layer [29] instead of the linear layer to help handle large intra-class variations of pulmonary nodules including variations in the nodule heterogeneity, size and shape [30]. Our new Maxout Local-Global Network is depicted in the Classification model in Figure 1.

### 3.3 Self-Training with Noisy Student model for lung cancer prediction

The main work in this paper is inspired by the original Self-training with Noisy Student paper [8]. In the paper, self-training is used to train a student model using 3 steps: (1) a teacher model is first trained on labelled images, (2) the teacher model is used to infer pseudo labels on unlabelled images and (3) the student model is then trained on both the labelled and pseudo-labelled images to obtain better performance. This process is repeated iteratively, whereby the student model is taken to be the new teacher to train a new student model in several iterative steps. In the process of training the student model, noise is added to the pseudo labels and the network to force the student to learn better from the combined labelled and pseudo-labelled dataset. Additionally, the student model is made to be bigger or at least as big as the teacher model as studies have shown that bigger models can learn better from bigger datasets [8,18,31]. The Self-training paper achieved state-of-the-art performance on the ImageNet dataset with a top-1 accuracy of 88.4% [8].

We adapted the Self-training with Noisy Student model for lung cancer prediction on the NLST dataset. In the context of lung cancer prediction on NLST, we first train our state-of-the-art Local-Global Network [9] teacher model to classify lung nodules on the labelled/annotated LIDC-IDRI dataset. Then, we use the teacher model to infer pseudo labels on 2,005 unlabelled images from the NLST dataset. From here onwards, we introduce a Maxout Layer in the Local-Global Network which replaces the fully connected layer in the original model. We then train a student model on both the pseudo-labelled and labelled data, and the student model then acts as the new teacher. This process is repeated for several iterations. The block diagram of our Noisy Student model for lung cancer prediction on the NLST dataset is shown in Figure 2.

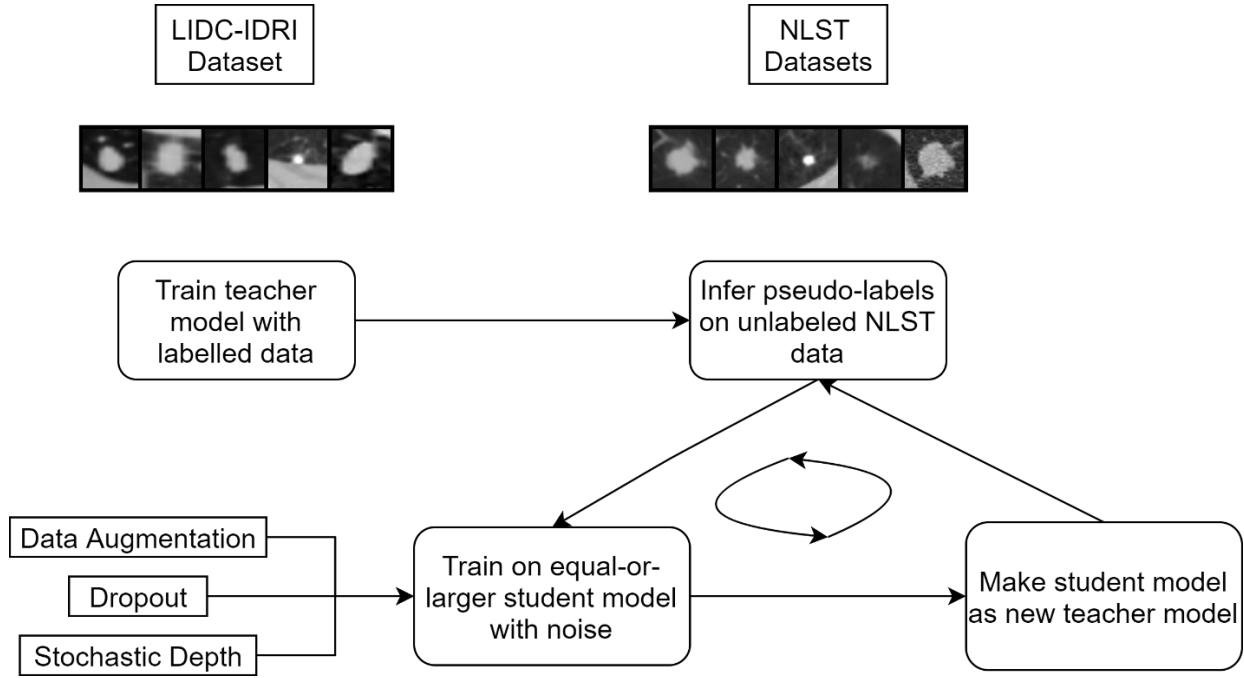

*Fig.2. Illustration of Noisy Student Training for Lung Cancer Prediction on the NLST dataset*

At the last iteration, the training is carried out with Maxout Local-Global Network and Mixup Maxout Local-Global Network separately. Finally, the nodule predictions for the NLST dataset are generated by the Mixup Maxout Local-Global classification network to calculate the probability of the patient having cancer or no cancer, using the Noisy-or method [32]. The Noisy-or method assumes every nodule is an independent cause of lung cancer and does not neglect the interaction of nodules within a patient [33]. Using the Noisy-or method, the probability of cancer is computed as:

$$P = 1 - \prod_i (1-P_i) \tag{1}$$

where $P_i$ represents the cancer probability of the $i$-th nodule.

In our Noisy Student framework, we incorporated the three sources of noise recommended by the original self-training paper [8]. The first two sources of noise are called "model noise", namely Stochastic depth and Dropouts. These model noise are used to introduce variations to the network architecture by dropping layers and nodes, respectively, so that new paths are learnt in the network training and to reduce the number of parameters (i.e., weights) that need to be fine-tuned by the network. The third source of noise is called "input noise", which is the common data augmentation procedure so that the student is forced to learn harder from images that are augmented from the original input image. We explain how the three types of noise are implemented in the subsections below.

### 3.3.1 Stochastic depth

Our self-training model applies stochastic depth as proposed by Xie et al. [8]. However, stochastic depth applied in our model produces different results from the original paper [8], as the size of our models for this application is rather small as compared to the bigger classification models that were used by the authors to predict larger datasets like ImageNet. To allow implementation of stochastic depth in our model, we separated the final layer and implemented stochastic depth with a survival probability $p_1$ of 0.8.

### 3.3.2 Dropout

In the Noisy Student Training paper by Xie et al. [8], dropouts were implemented in the model to serve as model noise and increase the performance of the model [34]. In our original Local-Global Network model, we had already implemented dropouts after each Non-Local block [9]. To analyze the role of dropouts further, in this study we included experiments to observe the performance of the model with and without any dropouts (see Appendix B.4).

### 3.3.3 Data augmentation

In the original Noisy Student Training paper by Xie et al. [8], extensive data augmentations were recommended to ensure that the student model is able to predict the same labels in the presence of slight translations in the input image, etc. This ensures the prediction consistency and robustness of the student model in the presence of different types of image noise and enables the student model to generalize to unseen/independent datasets.

Therefore, in line with the original Noisy Student Training paper, in our self-training model, we performed extensive data augmentations to help the student model learn better. We first cropped the nodules that were detected by DeepSEED. Then, from the axial, sagittal and coronal views of the nodule patches, we augmented each view with 9 different types of implemented augmentations altogether. The details of the 9 different augmentations are given in Appendix A.3 Self-training module.

### 3.4 Mixup regularization method

The Self-training and Noisy Student learning model in Section 3.3 was trained on pseudo-labelled images that might have been erroneous or corrupted as our pre-trained Local-Global Network classification model [9] is not completely error free. Data augmentation can help to prevent memorization of corrupt labels by our self-training model; however, this can be improved further by implementing Mixup [35] as a regularization technique.

Thus, we implemented the Mixup [35] method to further improve our model's robustness and generalization ability on the NLST dataset. In a nutshell, Mixup [35] is a technique that regularizes a classifier to behave linearly in between training examples. This prevents the model from memorizing corrupt labels and increases its robustness to adversarial examples. The proposed Mixup generic vicinal distribution equation is as follows:

$$\mu(\tilde{x}, \tilde{y} \mid x_i, y_i) = \frac{1}{n} \sum_j^n \mathbb{E}_\lambda [\delta(\tilde{x} = \lambda \cdot x_i + (1 - \lambda) \cdot x_j, \tilde{y} = \lambda \cdot y_i + (1 - \lambda) \cdot y_j)] \quad (2)$$

where $\lambda \sim Beta(\alpha, \alpha)$ for $\alpha \in (0, \infty)$; $(\tilde{x}, \tilde{y})$ is the virtual feature-target pair, $(x_i, y_i)$ and $(x_j, y_j)$ are two feature-target pairs drawn at random from the training set, and $\delta(\tilde{x} = \lambda \cdot x_i + (1 - \lambda) \cdot$

$x_j, \tilde{y} = \lambda \cdot y_i + (1 - \lambda) \cdot y_j)$ is a Dirac mass centred at $(\tilde{x}, \tilde{y})$. By sampling from the Mixup generic vicinal distribution, we obtain virtual feature-target vectors as follows:

$$\tilde{x} = \lambda x_i + (1 - \lambda) x_j \quad (3)$$

$$\tilde{y} = \lambda y_i + (1 - \lambda) y_j \quad (4)$$

where $\lambda \in [0,1]$. Varying the hyperparameter $\alpha$ helps to control the strength of interpolation between feature-target pairs. The $\alpha$ value is determined empirically, to determine the optimal interpolation strength between the feature-target pairs [35].

## 4. Experimental setup

In this section, we describe our experimental design and evaluation methods in detail. The training details of the nodule detection, classification and self-training networks are provided in Appendix A. All ablation studies are provided in Appendix B.

### 4.1 Datasets

In this study, we have utilised three different datasets: LUNA16, LIDC-IDRI and NLST. The CT number/distribution across the three datasets are summarized in Table 1.

*Table 1: Distribution of CT scans/number of patients across the three datasets used in this study.*

| Dataset | CT Number |
|---|---|
| NLST | 2005 |
| LUNA16 | 888 |
| LIDC-IDRI | 1018 |

#### 4.1.1 Marked/Annotated datasets

The two annotated datasets that we use in our study are the LUNA16 [27] and LIDC-IDRI [5] datasets. Specifically, these two datasets contain nodules that have been individually marked/annotated by 4 experienced thoracic radiologists. The LUNA16 dataset is based on the LUNA16 challenge dataset, which has 1186 nodules in total from 888 CT scans/patients. In the LIDC-IDRI dataset, four radiologists rated each nodule on a scale of one to five, where one and five, respectively, indicate that a nodule is clearly benign and clearly cancerous/malignant. Therefore, to aggregate all the nodule annotations by the four radiologists, similar to our previous studies [9,15,36], we took the median of the malignancy ratings as the ground truth and excluded nodules with a median rating of 3.

#### 4.1.2 Unmarked/Unannotated dataset

The National Lung Screening Trial (NLST) [7] dataset is another dataset that we used in this study. This dataset contains many scans (i.e., 53,454 patients altogether) and there are no detailed annotations for the nodule locations in the scans. Only a final diagnosis of whether a patient has cancer or is cancer-free is provided. We randomly selected 2,005 CT scans from this dataset for our study according to our available computational resources: 500 patients with benign nodules, 500 patients who have no cancer but with abnormalities, 402 patients with no cancer and without abnormalities and 603 patients with cancer. We will release the full list of patient IDs of the 2,005 scans on Github[1] upon acceptance of this manuscript.

## 4.2 Preprocessing

The steps for data preprocessing of each of the utilised datasets differ slightly, as we implemented the original methods proposed by the authors to ensure consistent performance from the models. The dataset we used for training the detection model is based on the LUNA16 dataset, which had its raw values clipped into the range [-1200,600] by the LUNA16 challenge organizers. After clipping the values, we transform the range linearly into [0,1] according to standard practices for normalizing datasets for deep learning [27]. We preprocessed the NLST dataset in a similar way to the LUNA16 dataset. For the LIDC-IDRI dataset, for training the classification model, we did not change any parameters and trained the model using the same method described in our previous publication [9]. We first clipped the values into the range [-1000,400] and normalized them to [0,1] after that.

## 4.3 Evaluation

To evaluate the performance of our proposed model, we compared the predicted labels with the ground-truth labels by computing four performance measures as follows:

| | |
|---|---|
| $$True\ positive\ rate, t_{pr} = \frac{TP}{TP + FN}$$ | (5) |
| $$Specificity = \frac{TN}{TN + FP}$$ | (6) |
| $$False\ positive\ rate, f_{pr} = 1 - Specificity = \frac{FP}{FP + TN}$$ | (7) |
| $$AUC = \int_0^1 t_{pr}(f_{pr}) df_{pr}$$ | (8) |

where $TP, TN, FN$, and $FP$ represent the number of true positives, true negatives, false negatives, and false positives, respectively. To compute the AUC, the false positive rate (i.e., $f_{pr}$) axis ranges from 0 to 1 in the ROC curve, and the true positive rate (i.e., $t_{pr}$), which is a function of the $f_{pr}$ axis also ranges from 0 to 1.

---

[1] *https://github.com/shakjm/rsslcp*

## 5. Results
### *5.1 Comparison of the proposed method with state-of-the-art methods*

In this section, we compare our proposed method with other state-of-the-art methods in the literature on the NLST dataset and tabulate the results in Table 2. Our original pre-trained model (i.e., Local-Global Network) [9] achieved AUC = 0.82, whereas Local-Global with the Maxout layer but without mixup regularization achieved AUC = 0.86. DeLong's test showed statistical significance at the 5% significance level when comparing the AUC value of the Mixup Maxout Local-Global network to both the Maxout Local-Global Network ($p = 0.0001$) and the Local-Global Network ($p < 2.2 \times 10^{-16}$). The performance of the Maxout Local-Global Network was also significantly different from the Local-Global Network ($p = 4.09 \times 10^{-10}$).

*Table 2: Comparison of our proposed Mixup Maxout Local-Global Network method with other state-of-the-art methods on the NLST dataset in the literature.*

| Proposed Method | Number of NLST CT Scans used for evaluation (in the testing set) | Obtained AUC result |
|---|---|---|
| Distanced LSTM [37] | 1794 | 0.83 |
| Quantitative Imaging Features [38] | 235 | 0.85 |
| Leaky Noisy-or Network [33] | 198 | 0.87 |
| DeepScreener [39] | 1359 | 0.86 |
| Local-Global Network (Ours) | 2005 | 0.82 |
| Maxout Local-Global Network (Ours) | 2005 | 0.86 |
| **Mixup Maxout Local-Global Network (Ours)** | **2005** | **0.87** |

Figure 4 shows the corresponding ROC curves of our three methods. From the ROC curves in Figure 4, we observe that our Mixup Maxout Local-Global network outperforms the other two methods across almost all FP rates. This demonstrates that the Mixup regularization technique worked effectively to improve the robustness of the Maxout Local-Global network and enabled it to generalize well on the independent 2,005 NLST scans used in this study.

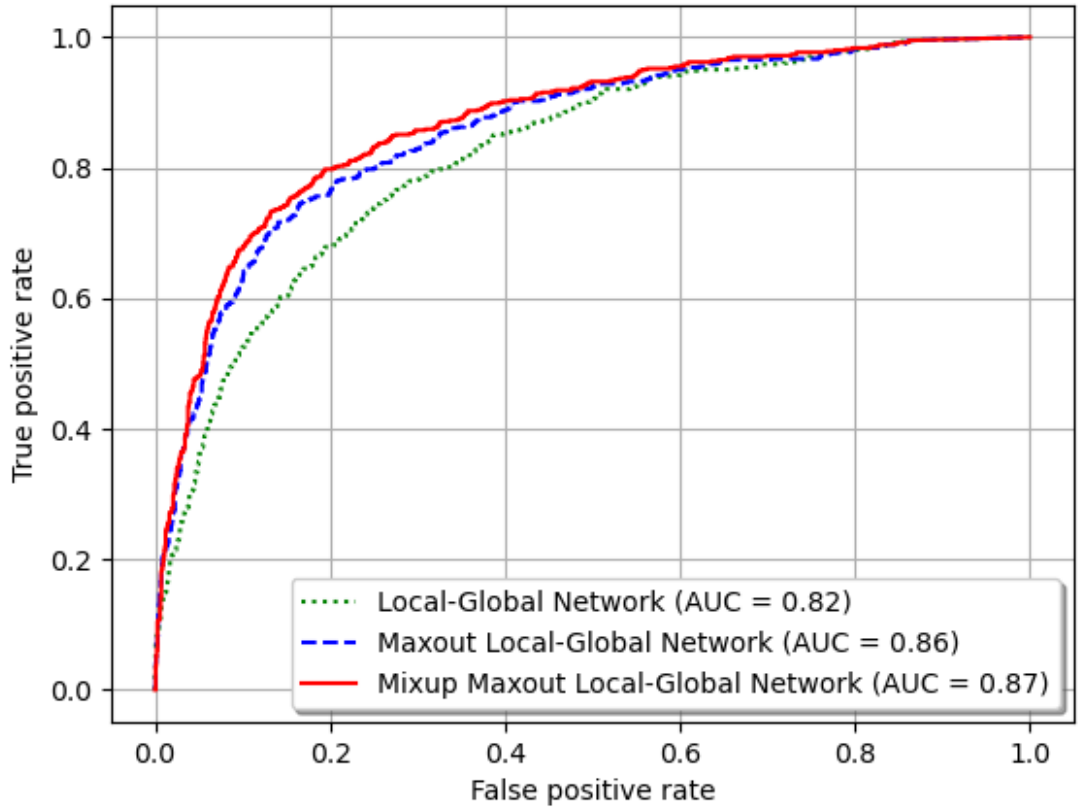

*Fig. 3.* *Comparisons of three receiver operating characteristic (ROC) curves corresponding to our new Mixup Maxout Local-Global Network with self-training, Maxout Local-Global Network with self-training, and Local-Global Network on 2,005 independent CT scans from the NLST dataset.*

Our model also outperforms all other proposed models in terms of the achieved AUC result and/or number of NLST scans used in the study. For example, we compare our results to that of the winner [33] of the Kaggle Data Science Bowl (DSB) 2017 challenge [40] – the testing set of the Kaggle DSB challenge consisted of 198 CT scans from the NLST dataset to evaluate the competitors' performance. Furthermore, the Kaggle DSB [40] training set consisted of NLST scans that were used by the Kaggle winner for training their method [33]. As compared to the Kaggle DSB winner [33], we evaluated our method on a much bigger dataset and obtained the same AUC result of 0.87.

Theoretically, training and testing a model on the same dataset would help the model familiarize itself to the nodule distribution of that dataset. Once the model is familiarized with the distribution of nodules and/or CT scan parameters of the training set, it would usually be able to generalize and perform better on the testing set [9,15,16]. However, in our experiments, we only trained our method on the LIDC-IDRI dataset and tested it on the NLST dataset, in the first teacher model. With this implementation, the data distribution for both training and testing sets are different. Our experimental setup simulates the setup/practice in the clinical environment and shows that our method can generalize to datasets in the screening environment, such as NLST, for accurately predicting lung cancer. The results show that our method outperforms or achieves at least the same AUC result as the other methods, whilst generalizing on a much bigger dataset of 2,005 scans.

## 6. Discussion

This study presents a new method for lung cancer prediction with several unique characteristics. Initially, we trained our first teacher model on the LIDC-IDRI dataset and tested it on the completely unseen NLST dataset with a different patient distribution and CT scan parameters. In contrast, other methods in the literature trained and tested their methods on the same dataset with the same patient distribution and CT parameters. For example, Balagurunathan et al. [14] trained their model on 244 NLST scans with patient data incorporated and evaluated their method on 235 scans. Theoretically, training and testing a model on the same dataset would help the model familiarize itself to the nodule distribution and CT parameters of the training set, which would enable it to generalize and perform better on the testing set [12,13,30]. The results show that our method outperforms or achieves at least the same AUC result as the other methods on a completely independent/unseen dataset, which mimics the real-world environment.

Second, we used a new Self-training with Noisy Student model to leverage on unannotated data from the large NLST dataset. The NLST dataset presents a vast resource of lung screening scans to validate/compare different deep learning methods; the nodules are unannotated, but the final diagnosis can still be used to evaluate different methods. To improve our classification performance, we employed a new Self-training with Noisy Student model to infer nodule predictions on the NLST dataset. By adding noise to the student model, it is forced to learn better from the large pseudo-labelled dataset. However, Self-training may not always improve the performance of the model, as distribution of dataset may affect how the model performs.

Third, we realize that the pseudo-labelled data might be erroneous as the pre-trained model that inferred the pseudo labels, although state-of-the-art, is not perfect and might generate corrupted predictions. Consequently, the student model will not be trained well, which leads to poor performance. Therefore, we used Mixup regularization to increase the robustness and generalizability of the student model. Furthermore, Mixup has been shown to prevent memorization of corrupt labels and increase robustness to adversarial examples – these are all crucial for applying our model to the completely independent NLST dataset.

A shortcoming of previous methods is that many of them have been evaluated on relatively small datasets, namely LUNA16 or LIDC-IDRI, for nodule detection and classification, respectively. Studies show that methods that have been trained on smaller datasets cannot be generalized to bigger screening datasets [19]. In contrast, our method achieved state-of-the-art results on a much bigger dataset of 2,005 independent screening CT scans.

Lastly, almost all previous methods have focused on the nodule detection or classification task individually. In this study, we trained, developed, and evaluated our new end-to-end scheme on a big dataset of around 4,000 CT scans altogether and achieved state-of-the-art results. The conventional CAD schemes produce too many false positives and are too distracting to radiologists. Conversely, our scheme could be used in clinical practice upon further evaluation on bigger datasets, which is in our plan for future work.

## 7. Conclusions

In this study, we presented a new Self-training with Noisy Student model for lung cancer prediction and obtained state-of-the-art results on the NLST dataset. We developed a new automatic end-to-end scheme that leverages on the vast resource of unlabelled nodules in NLST and showed that the results generalize to big independent datasets. To increase the robustness of our scheme to corrupt labels, we used Mixup regularization, and implemented a Maxout layer to handle large nodule variations. On a completely independent dataset of 2,005 scans, we achieved state-of-the-art performance even with a larger number of images as compared to other methods.

## 8. Acknowledgements

This work was supported by the Fundamental Research Grant Scheme (FRGS), Ministry of Education Malaysia (MOE), under grant FRGS/1/2018/ICT02/MUSM/03/1. This work was also supported by the Advanced Engineering Platform, Monash University Malaysia, and the TWAS-COMSTECH Joint Research Grant, UNESCO.

# Appendix

## A. Experimental implementation details

### A.1 Nodule detection network

We performed our simulations using the same parameters and implementation as the original DeepSEED paper [11], that is, we trained our model on the LUNA16 dataset using a ten-fold cross-validation method. We used Stochastic Gradient Descent (SGD) optimization and trained the model for 250 epochs. We also used a learning rate of 0.01 for 50 epochs, and then decayed the learning rate by 0.1 for every 50 epochs until 150 epochs was reached. From 150 epochs onwards, the learning rate remained constant until the end of the training procedure. We followed the parameters of the original model as closely as possible, except we used a batch size of 4 due to limitation in memory resources (as we used 2 2080Ti GPUs instead of 8 1080Ti GPUs used by the authors in the original study). We used the same cubic patch sizes that were recommended in the original study, which are 128 x 128 x 128 for training and 208 x 208 x 208 for testing.

### A.2 Nodule classification network

We reused all the parameters and implementation in our previous publication [9] and the reader is referred to our previous publication for their detailed description. In a nutshell, we initially trained our classification model using ten-fold cross-validation on the LIDC-IDRI dataset on a batch size of 256 for 250 epochs with a learning rate of 0.01. This was used to generate nodule predictions for the NLST dataset which are used as pseudo-labels for self-training. Similar to Kaggle DSB 2017 winner's settings [33], this model takes in a maximum of 5 nodules per patient (i.e., output by the nodule detection network) and generates a prediction for each of the nodule detections. We then calculate the probability of the patient having cancer by using the Noisy-or method in equation (1). In this way, we generated cancer/cancer-free predictions for all 2,005 patients in the NLST dataset.

### A.3 Self-training module

Noise is introduced in the models to ensure that the model learns well from the pseudo-labelled dataset. Similar to the Self-training with Noisy Student method [8], we carried out our experimentations with 3 types of noise, namely stochastic depth, dropout, and data augmentation. For stochastic depth, the survival probability of the final layer was set to 0.8. We applied dropouts of 0.1 and 0.2 after the first and second Non-Local blocks, respectively, in the pre-trained/original Local-Global network for lung nodule classification (see Figure 1). The following data augmentations were performed: 3 random rotations from $-180°$ to $180°$, 3 random additive Gaussian noise with the random rotations, and 3 different scales of Gaussian blurring with the random rotations. From these augmentations, we generated $3 \times 3 \times 3 = 27$ nodule crops for each detected nodule. The other self-training technique that we used in our study, which is similar to the methods prescribed in the original Self-training with Noisy Student paper [8], was using pseudo-labelled images with high confidence to train the new student models.

## B. Ablation studies

In this section, we perform our ablation studies to investigate the self-training techniques presented by Xie et al. [8] for predicting cancer/cancer-free patients. We performed all the ablation studies prescribed in [8] on our Maxout Local-Global Network. Additionally, we investigated the role of the Maxout layer on the network's performance. Finally, we analyzed varying the $\alpha$ parameter of Mixup on Maxout Local-Global Network's performance.

### B.1  Effects of implementing larger student models

We carried out various experimentations by implementing larger student models. These are performed to investigate whether larger student models help in improving the final AUC score. In our experiments, we modified both the Local-Global network and basic ResNet network. ResNet [41] blocks are known to perform well with depth increments and are known to be easily optimizable. This influenced our decision in implementing the simple ResNet architecture for our classification task. In this experiment, we have two main models, namely an extended Maxout Local-Global Network (i.e., Figure B.1) and an extended ResNet model (i.e., Figure B.2). To avoid confusion to the reader, we identify these corresponding models as Maxout A and ResNet A, which are larger than our proposed final model, Maxout Local-Global Network (cf. Figure 1). We have also implemented a Maxout layer in a separate ResNet A model (i.e., Figure B.3) to compare the performances of the proposed larger sized models. The results are tabulated in Table B.1.

*Table B. 1: Effects of implementing larger/different student models on self-training results*

| Method | AUC |
| --- | --- |
| Maxout A | 0.82 |
| ResNet A | 0.85 |
| ResNet A with Maxout layer | 0.84 |
| Maxout Local-Global Network | 0.86 |

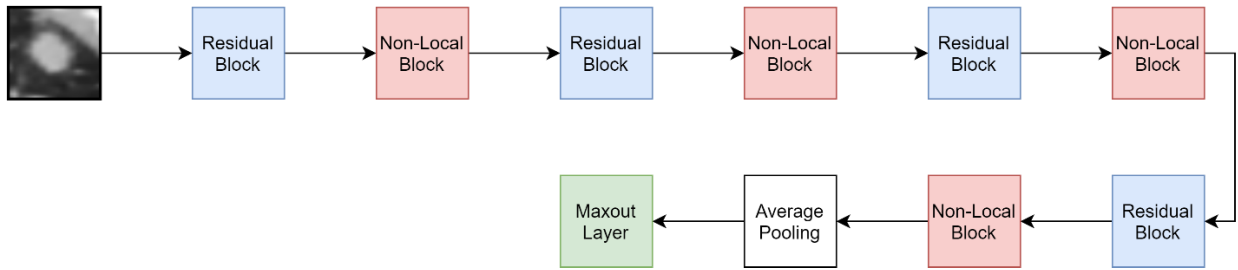

***Fig B. 1.** Architecture of Maxout A Network.*

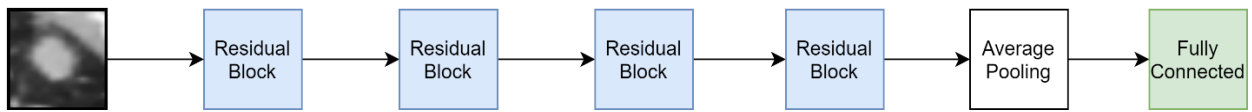

***Fig B. 2.** Architecture of ResNet A Network.*

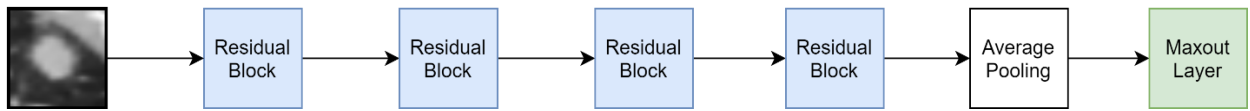

*Fig B. 3. Architecture of ResNet A with Maxout Layer Network.*

From the results, we observe that the proposed Maxout Local-Global network still outperforms all the other models. Having larger models might require more parameter tuning to refine/tune the additional parameters in these models. We conclude that larger models do not necessarily produce better results, especially with relatively smaller datasets as compared to ImageNet.

### B.2 Hard pseudo Labels vs soft pseudo labels

We further performed experimentations on training the student models on hard pseudo labels versus soft pseudo labels and compared their performance. The experiments were carried out on our Maxout Local-Global Network; all parameters/settings were identical apart from the hard/soft pseudo labels that were varied in this ablation study. The results are tabulated in Table B.2.

*Table B. 2: Effects of implementing soft pseudo labels versus hard pseudo labels in training the student models*

| Method | AUC |
|---|---|
| Hard labels | 0.82 |
| Soft labels | 0.86 |

From the results, we observe that training our student models with soft pseudo-labels produced better results than hard ones. In the original self-training paper [8], the authors concluded that using soft pseudo-labels or hard pseudo-labels must be determined on a case-by-case basis. To the best of our knowledge, this is the first time that this study is being performed for lung cancer prediction and our experiments show that using soft pseudo-labels might be better for this application.

### B.3 Effects of warm-starting the student model

The student model goes through warm-starting by partially training the teacher model, and then using the partially trained teacher model as a student model, and continue training the student model with the pseudo labels. Warm starting the student model might produce better results than training a student model from scratch. To analyze this hypothesis, we performed self-training with and without warm-starting the student model and tabulated the results in Table B.3.

*Table B. 3: Effects of warm-starting the student model*

| Method | AUC |
|---|---|
| Maxout Local-Global Network w. warm-starting | 0.81 |
| Maxout Local-Global Network w/o warm-starting | 0.86 |

From the results, we observe training the student model from scratch is better than warm-starting the student model. The obtained results are similar to Xie et al. [8] and confirm that warm-starting the student model might lead to sub-optimal results as the student initialized with the teacher can get stuck in a local optimum. Therefore, we conclude that training the student model from scratch is better than warm-starting the student model in our study.

### B.4 Importance of noise in self-training

In this ablation study, we perform similar studies to Xie et al. [8] and analyze the effects of noise on the unlabelled data/NLST dataset. To do this, we gradually removed the augmentation, stochastic depth and dropout for the unlabelled data when training the student model. We also examined the effects of having a noised teacher versus an un-noised teacher to study whether it is necessary to disable noise in the teacher model whilst generating the pseudo labels. All the results are tabulated in Table B.4.

*Table B. 4: Ablation study of noising as prescribed by the Self-training with Noisy student paper [8]. Three sources of noise were recommended in [8] and examined here: (1) Data augmentation (Aug), (2) Stochastic depth (SD) and (3) Dropout. Our original pre-trained Local-Global Network was used as the teacher model and used for the noisy student training.*

| Model | AUC |
|---|---|
| Local-Global Network | 0.82 |
| Noisy Student Training | 0.78 |
|     student w/o. Aug | 0.83 |
|     student w/o. Aug, w/o. SD | 0.86 |
|     student w/o. Aug, w/o. SD, w/o. Dropout | 0.82 |
|     teacher w. Aug, w. SD, w. Dropout | 0.78 |

From Table B.4, we observe that adding noise in the form of dropouts is important in helping the model learn better (AUC = 0.86). With stochastic depth and dropout implemented, there is a small increment over the original Local-Global network (AUC = 0.83). When no noise is introduced, the model performs similarly to the original Local-Global network without noise (AUC = 0.82). As our model size is comparatively smaller, stochastic depth and/or data augmentation may not be beneficial for performance improvement. For example, in the paper proposed by Xie et al. [8], the models that are implemented are much larger in size and the datasets used are many times bigger than the NLST dataset we have utilized here. Thus, having dropouts in our model is sufficient to improve the self-training network. Finally, an un-noised teacher model produces better performance than a noised one, which is a similar result as that obtained in [8]. This result demonstrates that it is important to have a powerful teacher model to generate accurate pseudo labels, as the better the performance of the teacher, the better the student can learn from the teacher, which leads to higher performance by the student.

### B.5 Maxout layer implementation

Our original Local-Global Network [9] implemented a linear layer instead of Maxout as the final layer. We implemented two different models to analyze the effects of having Maxout replacing the original final linear layer and tabulated the results in Table B.5.

*Table B. 5: Comparison of results between using a Linear layer versus a Maxout layer in our Local-Global Network*

| Method | AUC |
| --- | --- |
| Local-Global Network | 0.82 |
| Maxout Local-Global Network | 0.86 |

From Table B.5, we observe that replacing the linear layer with a Maxout layer produces a better AUC result. Implementing the Maxout layer was shown to improve the performance of a lung nodule detection scheme as it handles variations of lung nodules well [30]. As the appearance of lung nodules vary greatly in terms of their texture, size, lobulation, spiculation, etc. and these characteristics are all important to predict lung cancer occurrence [42], it is highly important to handle nodule variations for our task of predicting lung cancer occurrence. The results show that implementing a Maxout layer is also important for the classification task.

### B.6 Mixup regularization experiments

We performed a study of varying the Mixup hyper-parameter, $\alpha$ on the results. Mixup regularization prevents the model from memorizing the training set, which can be regulated by varying the value of the $\alpha$ hyper-parameter [35]. Following the authors of the original Mixup paper [35], we varied the $\alpha$ value between 0.1 to 32 and the results are tabulated in Table B.6.

*Table B. 6: Mixup Maxout Local-Global Network experiments with different $\alpha$ values.*

| Method | AUC |
| --- | --- |
| Maxout Local-Global Network | 0.86 |
| Mixup Maxout Local-Global Network | |
| $\alpha = 0.1$ | 0.85 |
| $\alpha = 0.2$ | 0.86 |
| $\alpha = 0.4$ | 0.86 |
| $\alpha = 0.8$ | 0.86 |
| $\alpha = 1$ | 0.86 |
| $\alpha = 2$ | 0.86 |
| $\alpha = 4$ | 0.86 |
| $\alpha = 8$ | 0.86 |
| $\alpha = 16$ | 0.87 |
| $\alpha = 32$ | 0.87 |

From the results in Table B.6, we observe that the best AUC result is achieved with $\alpha = 16$. Incrementing $\alpha$ beyond 16 does not lead to further improvement in the results. We had also observed in the Results section that implementing Mixup significantly improved the AUC result, whereby the AUC value of the Mixup Maxout Local-Global network was statistically significantly better than the Maxout Local-Global Network using DeLong's test ($p = 0.0001$).

### B.7  Ablation Study Summary

From the results of our ablation studies, we can conclude that soft pseudo labels, not warm-starting the student model, and introducing noise in the form of dropouts improves the self-training performance. We also observed that implementing larger student models does not necessarily lead to better performance as more parameter tuning to refine/tune the additional parameters in these models may be required and our study was also performed on a relatively smaller dataset as compared to ImageNet. We also conclude that implementing the Maxout layer and Mixup regularization both help our model to distinguish between cancer/cancer-free cases better.